\documentclass[9pt,twocolumn,twoside]{osajnl}
\pdfoutput=1 
\journal{ol} 

\setboolean{shortarticle}{true}


\title{Tunable telecom to mid-infrared optical parametric oscillation via microring-based $\chi^{(3)}$ nonlinearities}

\author[1,$\#$,$\dag$]{Yulong Tang}
\author[1,$\#$]{Zheng Gong}
\author{Xiangwen Liu}
\author[1,*]{Hong X. Tang}

\affil[]{Department of Electrical Engineering, Yale University, New Haven, Connecticut 06520, USA}
\affil[*]{Corresponding author: hong.tang@yale.edu}
\affil[$\dag$]{Present address: School of Physics and Astronomy, Key Laboratory for Laser Plasmas (MOE), Collaborative Innovation Center of IFSA,Shanghai Jiao Tong University, Shanghai 200240, China; email:yulong@sjtu.edu.}
\affil[$\#$]{These authors contributed equally}

\ociscodes{(190.0190) Nonlinear optics; (190.4223) Nonlinear wave mixing; (140.3948) Microcavity devices.}

\begin{abstract}
Optical parametric oscillation (OPO) with far-shifted frequency sidebands has attracted significant interests in precision spectroscopy and quantum information processing. Microresonator based OPO sources hold the advantages of miniaturized footprint and versatile dispersion engineering. Here we demonstrate large-frequency-shifted $\chi^{(3)}$-based OPO from crystalline aluminum nitride microrings pumped at $\sim$2 $\mu$m in the normal dispersion regime. OPO in the telecom and mid-infrared bands with a frequency separation of 65.5 THz is achieved. The OPO frequency can be agilely tuned in the ranges of 10, 1 and 0.1 THz respectively by tailoring the microring dimensions, shifting the pump wavelength, and controlling the chip temperature. At high intracavity pump powers, the OPO sidebands further evolve into localized frequency comb lines. Such telecom to mid-infrared OPO with flexible wavelength tunability will lead to enhanced chip-scale light sources.   
\end{abstract}
\setboolean{displaycopyright}{true}
\begin{document}
\maketitle

\hfill\break
Broadband and widely-tunable coherent light sources are essential tools for various applications \cite{Duarte2008}. While the available emission spectra of conventional lasers are set by their gain media, nonlinear optical frequency conversions help generate coherent light ranging from ultraviolet to mid-infrared (mid-IR) regions \cite{Savage:10}. Among these nonlinear optical processes, $\chi^{(3)}$-based optical parametric oscillation (OPO) features the generation of coherent frequency components with two sidebands equally separated from the pump and has attracted significant interest for its high conversion efficiency and large sideband frequency tunability \cite{Srinivasan:16, Sayson:19}. Such OPOs have been extensively studied in silica fiber cavities  \cite{Wong:07,Bessin:17}, but with limited bandwidth due to inconvenient dispersion engineering and high pump power requirement. 

On the other hand, optical microresonators, possessing high quality-factors (Qs) as well as small mode volumes, have proven as excellent platforms for ultra-low threshold OPO generation \cite{Kippenberg:04, bruch2019chip}. Meanwhile, the versatile and precise dispersion engineering in microresonators could facilitate large-frequency-shifted or octave-spanning OPOs in the desired spectrum regions \cite{Sayson:19}. For instance, by tailoring higher-order dispersion (even orders) while pumping in the normal dispersion regime, OPO sidebands extending to mid-IR regimes of 2 and 2.4 um have been achieved in silica and magnesium fluoride microresonators with corresponding frequency shifts of 85 and 140 THz, respectively \cite{Sayson:17,Fujii:19}. Such broadband OPO sources could enable applications in chemical and molecule detection owing to fluent characteristic vibrational and rotational transitions of many important molecules lie in the mid-IR regime (2.5-20 µm) \cite{Lecaplain}.  

Apart from whispering gallery microcavities, nanophotonic microring resonators based on silicon nitride ($Si_3N_4$) platforms have also been employed to generate low threshold OPO \cite{Gaeta:17} in the telecom band \cite{Gaeta:19}, near infrared \cite{Srinivasan:16}, and even down to the visible wavelength region \cite{Srinivasan:19}. However, to the best of our knowledge, widely-tunable OPO sidebands that extend to the mid-IR regime have not been demonstrated in a nanophotonic platform. Recently, aluminum nitride (AlN) has emerged as an appealing material for integrated nonlinear photonics thanks to its strong quadratic ($\chi^{(2)}$) and cubic ($\chi^{(3)}$) susceptibilities \cite{Hong1,Hong2,Hong3}. Single-crystalline AlN, in particular, has shown improved device performance \cite{Liu587, Liu1279} and has stimulated progresses including frequency comb generation \cite{Gong4366,Liu1943}, second harmonic conversion \cite{Bruch113}, ultraviolet supercontinuum combs \cite{Liu2971} and efficient OPO sources \cite{bruch2019chip}. While the extent of these advances spans from the ultraviolet to near-infrared regimes, the broad transparency of AlN (from 200 nm to 10 $\mu$m) \cite{Hong1,Lin} also allows for nonlinear frequency conversion into the mid-IR region.

In this letter, we demonstrate large frequency-shifted $\chi^{(3)}$ OPOs in crystalline AlN microrings pumped at $\sim$2 $\mu$m. By tailoring the microring dimensions, desirable phase-match conditions can be obtained for OPO generation at the telecom and mid-IR regions with a frequency separation of 65.5 THz. We then show that the OPO frequency sidebands can be tuned by $\sim$10 THz over a varied microring width of 200 nm, which is consistent with our numerical simulations. Further fine-tuning of the OPO sideband frequencies is realized by varying the pump wavelength and controlling the chip temperature with tuning scales of $\sim$1 THz and $\sim$0.1 THz, respectively.

Our microring resonators are made out of a ~1-µm thick epitaxial AlN film grown on a c-plane sapphire substrate via metal organic chemical vapor deposition \cite{Yan103001}. The microring patterns are defined by 100 kV electron beam lithography and transferred into the AlN film using a two-step dry etching process \cite{Gong4366,Fan5850}. At last, the devices are encapsulated within $\sim$1.5-µm thick silicon dioxide (SiO$_2$) via plasma enhanced chemical vapor deposition. Figure 1(a) shows the scanning electron microscope (SEM) image of the fabricated straight waveguide-coupled microring resonator. And Fig. 1(b) schematically illustrates the OPO generation process in a Kerr microcavity, where the pump field inside the cavity gives rise to oscillations at two new frequency components based on $2\omega_p=\omega_s+\omega_i$. Such process can be enhanced if $\omega_s$ and $\omega_i$ overlap the microring resonances.

\begin{figure}[htbp]
    \centering
    \includegraphics[width=1\linewidth]{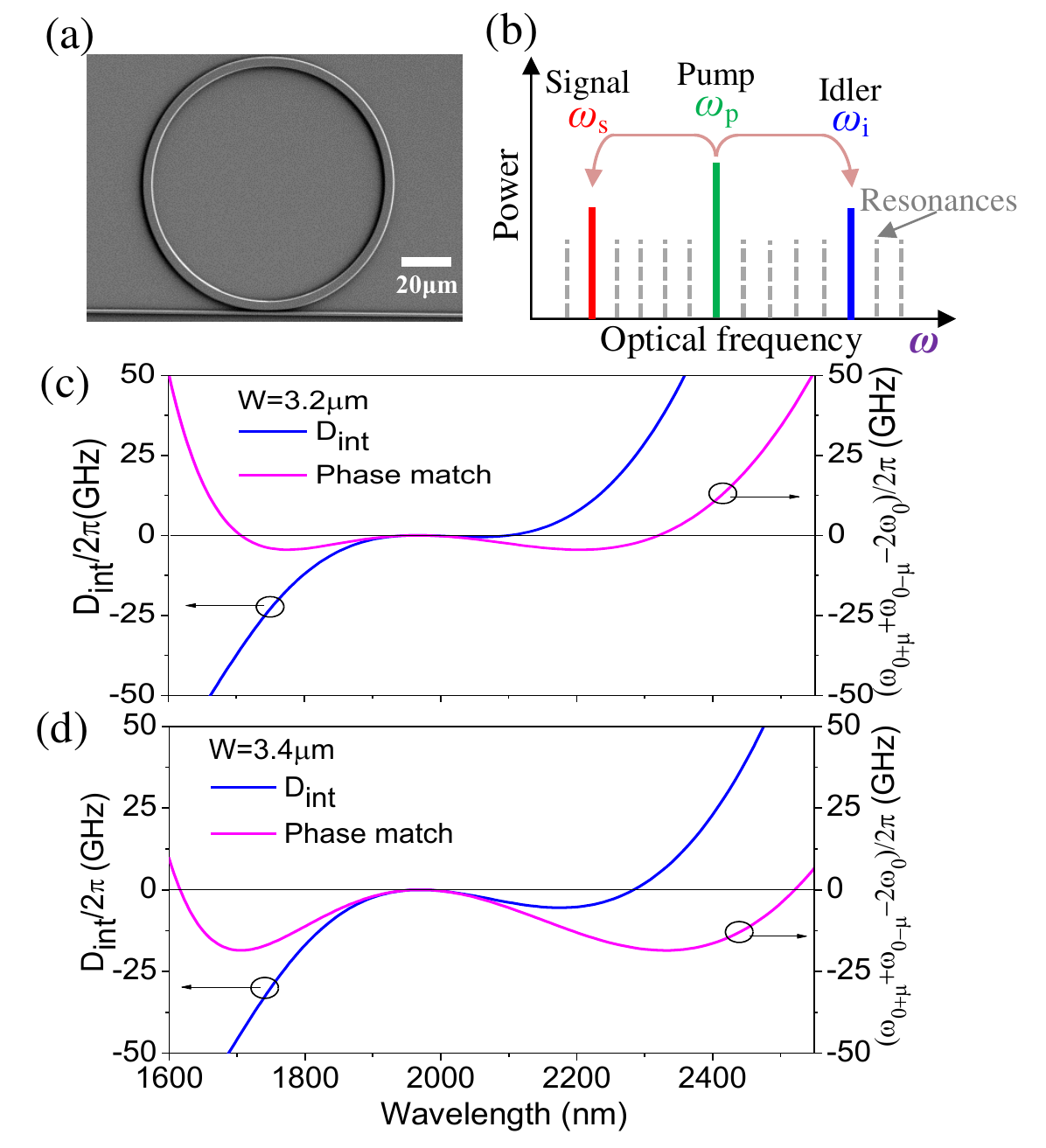}
    \caption{(a) SEM image of the AlN microring with a radius R$=$50 $\mu$m. (b) Schematic of the $\chi^{(3)}$-based OPO sideband generation in the microcavities with $\omega_p$, $\omega_s$ and $\omega_i$ respectively being the pump, signal and idler frequency. (c,d) Simulated TE$_{00}$–mode integrated dispersion (blue) and phase matching curves (magenta) for the microing width W$=$3.2, 3.4 $\mu$m and R$=$50 $\mu$m using the finite element method, where the crossing-zero points of the magenta curve indicate the phase-matched modes for effective OPO.}
    \label{fig.1}
\end{figure}

For the device design, we fix the microring radius at 50 µm while tailoring the microring width to adjust OPO phase-matching conditions.
After considering up to the $4^{th}$ order dispersion, the cavity modes frequencies can be expanded around the pump angular frequency $\omega_0$ as \cite{Fujii:19} $\omega_{\mu}=\omega_0+D_1\mu+(1/2!)D_2\mu^2+(1/3!)D_3\mu^3+(1/4!)D_4\mu^4$, where \textit{µ} is the mode number relative to the pump mode and $D_i$ denotes the $i^{th}$ order dispersion coefficient at $\omega_0$. Meanwhile, the OPO sidebands are determined by the phase-matching condition \cite{Fujii:19}: $\omega_{0+\mu}+\omega_{0-\mu}-2\omega_0=D_2\mu^2+(1/12)D_4\mu^4\simeq0$.
To achieve widely-separated OPOs, we tailor the microring geometries to have $D_2<0$ and $D_4>0$ so that the phase-matching curve of $(\omega_{0+\mu}+\omega_{0-\mu}-2\omega_0)/2\pi$ can be bent to have cross-zero points far away from the pump. Figures 1(c, d) show the simulated $D_{int}$ (blue) and phase-matching curves (magenta) for the TE$_{00}$ mode in the microrings with widths of W=3.2 and 3.4 $\mu$m, respectively. For the 3.2-µm-wide micoring, the phase matching points are at $\sim$1.71 and $\sim$2.31 $\mu$m, corresponding to $\mu\approx\pm52$ with $D_2/(2\pi)\approx$-12 MHz and $D_4/(2\pi)\approx$59 kHz; while the 3.4-µm-wide microring exhibits phase-matching points at $\sim$1.61 and $\sim$2.51 $\mu$m, corresponding to $\mu\approx\pm73$ with $D_2/(2\pi)\approx$-26 MHz and $D_4/(2\pi)\approx$57 kHz. The larger frequency-shifted phase-matching points from the pump in wider microrings is mainly due to the decrease of ${D_2}$ at similar ${D_4}$.

The experimental setup is schematically presented in Fig. 2(a). The pump source was a narrow-linewidth tunable seed laser (1940-1980 nm) amplified with a thulium-doped fiber amplifier (TDFA). A fiber-based polarization controller (FPC) is used to control the input light polarization before launching it into the waveguide with a mid-IR aspheric lens pair. And the output signal, collected with another aspheric lens pair, is sent into an optical spectrum analyzer (Yokogawa AQ6376, 1500-3400 µm).  The total transmission efficiency of the chip is $\sim$10$\%$. Figures 2(b-e) present the measured transmission spectra of the microrings (radius of R$=$50 $\mu$m, widths of W$=$3.2, 3.4 µm), where the free spectral range (FSR) is $\sim$450 GHz and the extracted loaded Q-factors are $\sim$480 k and $\sim$570 k around the pump wavelengths, respectively. 

Along with experimental investigation,  numerical simulations are also carried out to study the OPO process under the dispersion profiles in Figs.1(c, d) based on the coupled mode equations \cite{Herr:14}:  
\begin{multline}
\frac{\partial{A_\mu}}{\partial{t}}=-\frac{\kappa_0+\kappa_{ext}}{2}+\delta_{\mu_0}\sqrt{\kappa_{ext}}S_{in}e^{-i(\omega_p-\omega_0)t}+\\
ig\sum_{\mu^\prime,\mu^{\prime\prime},\mu^{\prime\prime\prime}}A_{\mu^\prime}A_{\mu^{\prime\prime}}A^\star_{\mu^{\prime\prime\prime}}e^{-i(\omega_{\mu^\prime}+\omega_{\mu^{\prime\prime}}-\omega_{\mu^{\prime\prime\prime}}-\omega_\mu)t}.     \hspace{1cm} 
\end{multline}   

Here, $A_\mu$ is the amplitude of the mode with a relative mode number \textit{µ}, $|{A_\mu}|^2$ describe the photon number of each mode, and $t$ is the time.  $\kappa_0$ and $\kappa_{ext}$ denote the intrinsic decay and coupling rates of the cavity, respectively. $|S_{in}|=\sqrt{P_{in}/\hbar\omega_0}$ is the amplitude of the pump power, and $\delta_{\mu_0}$ is the Kronecker delta. The summation term includes all $\mu^\prime,\mu^{\prime\prime},\mu^{\prime\prime\prime}$ satisfying the relation $\mu=\mu^\prime+\mu^{\prime\prime}-\mu^{\prime\prime\prime}$. The cubic Kerr nonlinearity of the system is described by the nonlinear coupling coefficient:
\begin{equation}
 g=\frac{\hbar{\omega_0}^2cn_2}{{n_0}^2V_{eff}},     \hspace{5.7cm}  
\end{equation}
where $n_0$ is the refractive index, $n_2$ is the Kerr nonlinear refractive index, \textit{c} is the light speed in vacuum, $\hbar$ is the reduced Planck constant, and $V_{eff}$ is the effective cavity mode volume. The amplitude of the output power from the cavity $|S_{out}|=\sqrt{P_{out}/\hbar\omega_0}$ can be obtained by:
\begin{equation}
 S_{out}=S_{in}-\sqrt{\kappa_{ext}}\sum{A_{\mu}e^{-i(\omega_{\mu}-\omega_p)t}}.  \hspace{2.5cm} 
\end{equation}
In simulation, similar parameters as in the experiment are adopted, and normalized pump-resonance detuning $2(\omega_{p}-\omega_{0})/(\kappa_{0}+\kappa_{ext})$ is scanned to trigger OPO.  

\begin{figure}[htbp]
    \centering
    \includegraphics[width=1\linewidth]{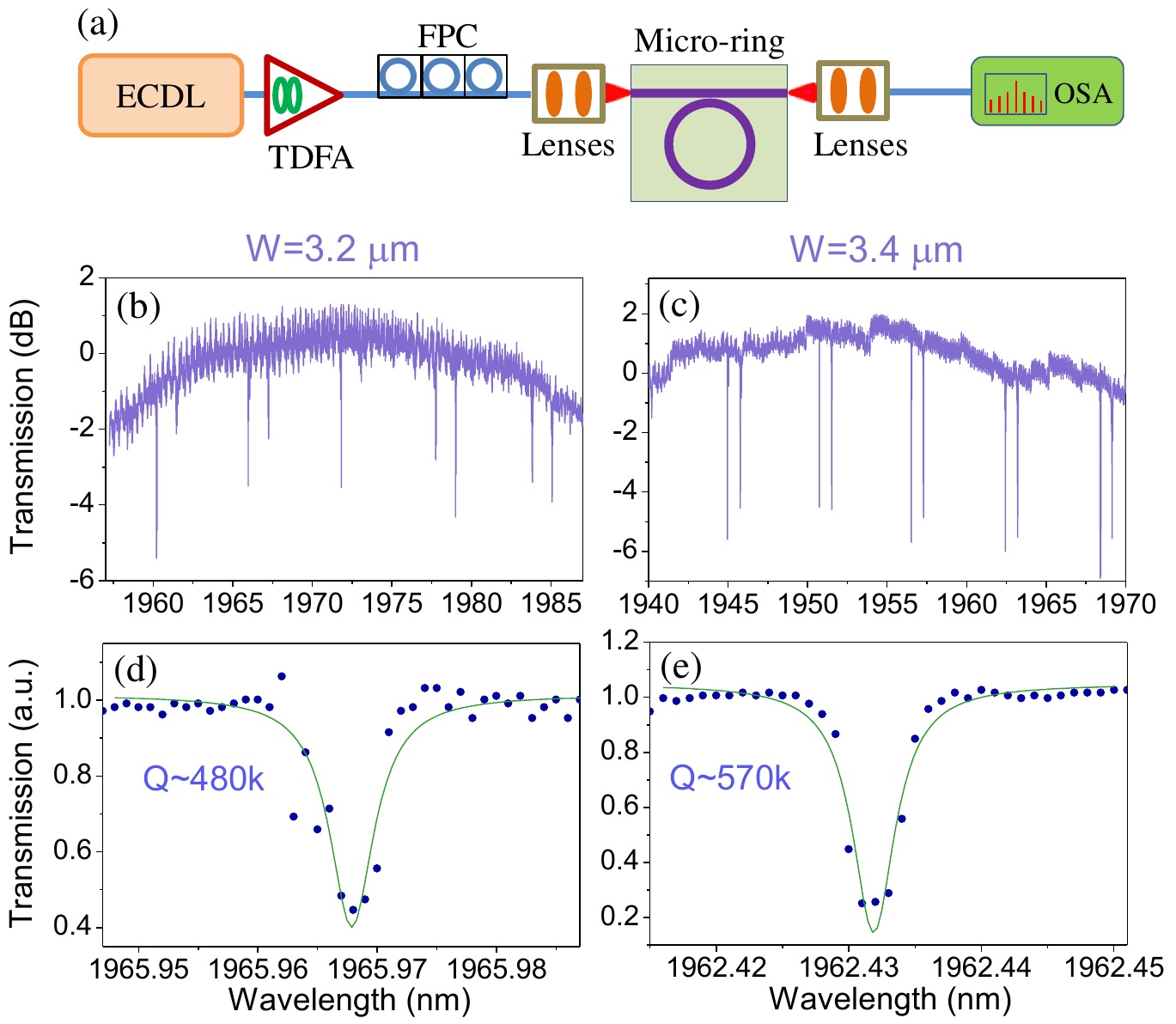}
    \caption{(a) The experimental setup. ECDL: external-cavity diode laser; TDFA: thulium-doped fiber amplifier; FPC: fiber polarization controller; OSA: optical spectrum analyzer. (b,c) TE$_{00}$ mode transmission spectra for microrings W$=$3.2, 3.4 $\mu$m and R$=$50 $\mu$m. (d,e) The zoom-in views of the resonances at $\sim$1965.97 and $\sim$1962.43 nm in (b,c) respectively; dots show the measured data while the curve is Lorentz fitting.}
    \label{fig.2}
\end{figure}

Figures 3(a, b) show the measured OPO spectra from the microrings with an on-chip pump power of ~375 mW. For the 3.2-µm-wide microring, the OPO sidebands show up at ~1711.6 and ~2312 nm, which agrees well with the cross-zero points of the phase-matching curve in Fig. 1(c), as well as the simulated sidebands in Fig. 3(c). The total frequency shift from the signal (at 2312 nm) to the idler (at 1711.6 nm) is over 45.5 THz (around 600 nm). Upon increasing the microring width to 3.4 $\mu$m, extended OPO frequency shift is obtained which is also consistent with the simulation in Fig.3 (d). In this case, the two OPO sidebands are observed to appear at 1618.2 and 2503 nm with a total frequency separation of over 65.5 THz (>880 nm), which is 20 THz broader than that in the 3.2-µm-wide device. These results indicate that frequency tuning of the OPO sidebands over $\sim$10 THz can be achieved in an AlN nanophotonic platform by adjusting the microring width of 200 nm. 

\begin{figure}[htbp]
    \centering
    \includegraphics[width=0.9\linewidth]{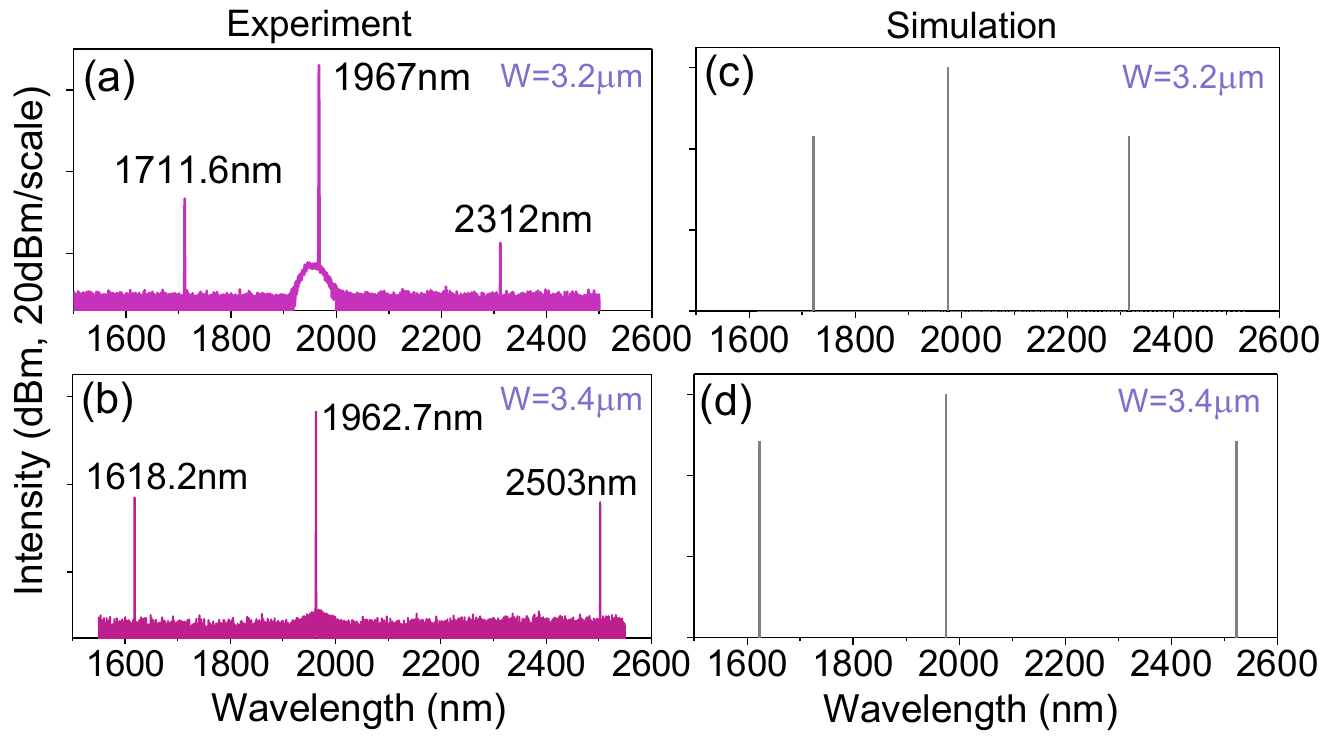}
    \caption{Measured OPO spectra from the microrings of R$=$50 $\mu$m, W$=$3.2 $\mu$m (a) and W$=$3.4 $\mu$m (b). (c,d) Simulated OPO spectra corresponding to (a,b) under the D$_{int}$ and phase-matching curves shown in Fig.1(c,d).} 
    \label{fig.3}
\end{figure}

Apart from controlling the dispersion profile via tailoring the microring geometries, the OPO sideband frequencies can also be tuned by varying the pump wavelength within the $D_2$<0 regime. Figure 4 illustrates how the OPO sidebands generate in the 3.4-µm-wide microring when blue-shifting the pump from one resonance to another. As can be seen, by decreasing the pump wavelength, the signal (idler) sideband shifts to the red (blue) spectral region, resulting in a widened spacing between the signal/idler modes. In this case, the signal (idler) frequency can be shifted by $\sim$3 ($\sim$2.5) THz in total as the pump is tuned by $\sim$2.1 THz. Such OPO frequencies tuning effect is also numerically simulated and plotted as the star dots in Fig. 4, which trend the experimental results. 

\begin{figure}[htbp]
\centering
\includegraphics[width=0.9\linewidth]{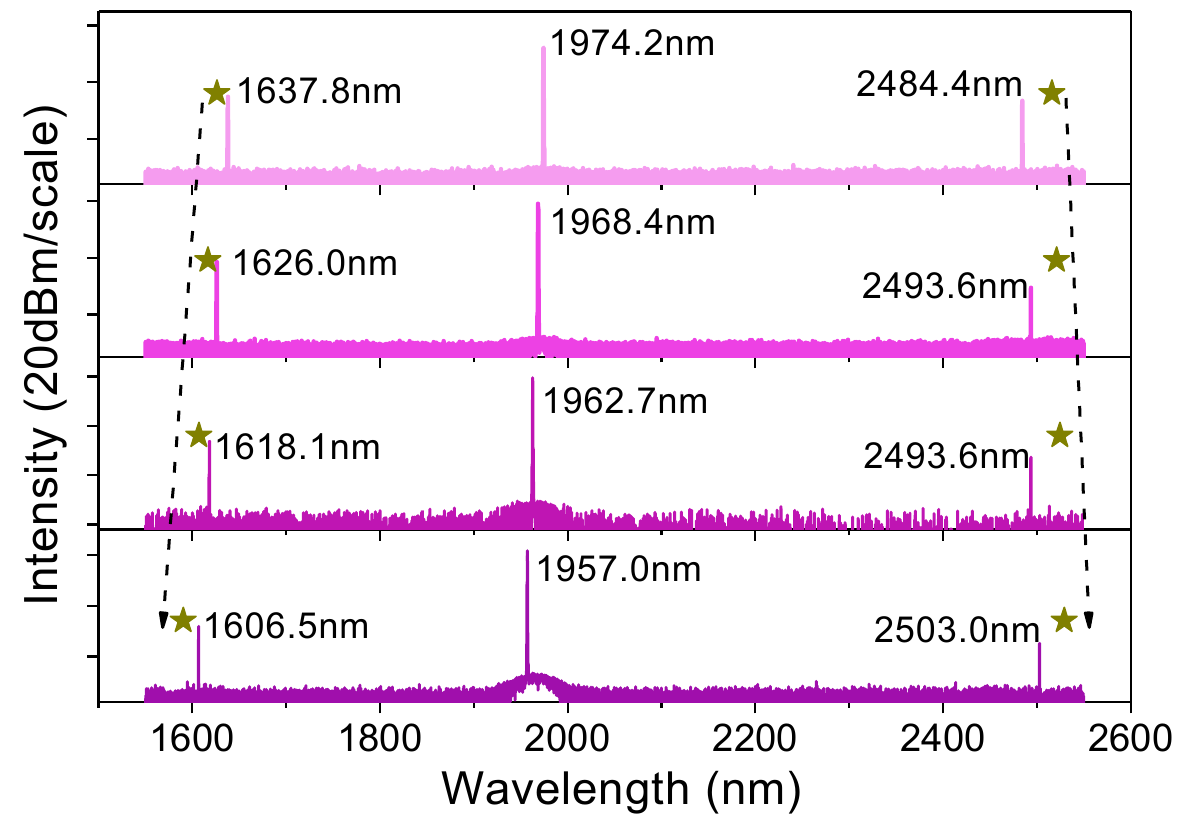}
    \caption{OPO sidebands tuning in the microring of W$=$3.4 $\mu$m and R$=$50 $\mu$m by tuning the pump laser to different resonances. Star dots show the simulated OPO sideband positions under the same pump wavelengths as the experiments.}
    \label{fig.4}
\end{figure}
\begin{figure}[htbp]
    \centering
    \includegraphics[width=0.9\linewidth]{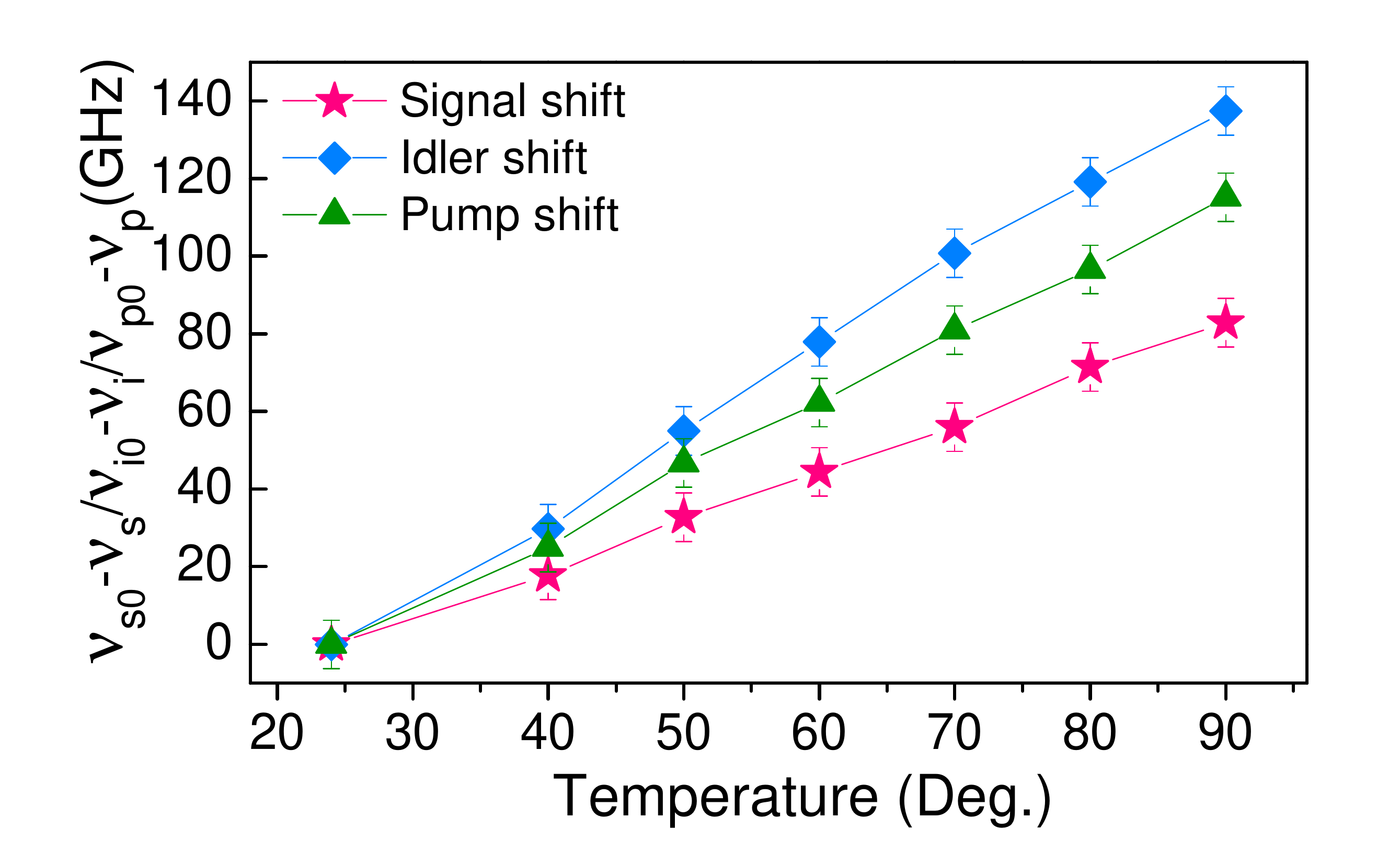}
    \caption{Temperature tuning of OPO signal and idler sidebands in the microring of W$=$3.4 $\mu$m and R$=$50 $\mu$m. The pump wavelength is adjusted accordingly as the temperature is varied. $\nu_{p0}$, $\nu_{i0}$ and $\nu_{s0}$ are the pump frequency and that of the two generated OPO sidebands when the chip's temperature is at 24 \textdegree{C}, corresponding to wavelengths of 1962.8, 1618.2, and 2493.8 nm. The error bar results from the OSA's resolution ($\sim$0.1 nm).}
    \label{fig.5}
\end{figure}

While the above OPO frequency tuning mechanisms (Figs. 3 and 4) provide discrete and relative large tuning steps, quasi-continuous frequency tuning can also be achieved by utilizing the thermal-optic effect in the AlN microrings \cite{Gong4366}. For this purpose, we mounted an external heater (precision of 0.5 \textdegree{C}) below the chip for effective temperature tuning  and recorded the temperature-dependent OPO spectra simultaneously. The observed tuning of the signal/idler frequencies under fixed pump power are summarized in Fig. 5. By varying the heater temperature from 24 to 90 \textdegree{C}, the signal and idler sidebands red-shift by 82.9 GHz (1.72 nm) and 137.4 GHz (1.2 nm), respectively. Or equivalently, the OPO sideband frequencies can be tuned at a step of $\sim$0.75 GHz (0.013 nm) for the signal and $\sim$1.25 GHz (0.009 nm) for the idler, which is limited by the 0.5 \textdegree{C} precision of the temperature controller and if desired can be further improved by embedding the device in a ovenized heater. 

By further tuning the pump into the resonances, single-FSR-spaced four-wave mixing (FWM) comb lines are subsequently generated around the pump and the primary OPO sidebands. These comb spectra are shown in Figs. 6(a,b). Note that the 3.2-$\mu$m-wide microring generates stronger comb lines that emerge above the OSA noise floor over a full 800~nm wavelength span, as shown in Fig. 6(a), while the comb lines from the 3.4-$\mu$m-wide microring (Fig. 6(b)) are clustered around the signal, pump and idler which are widely separated. Figures 6(c,d) are the simulated comb lines for the two microrings, showing good agreement with experimental observations. In our experiment and simulation, the involved comb lines are found to be in incoherent/noisy states. Similar localized frequency comb generation has been observed in crystalline MgF$_2$ microresonators \cite{Sayson:18}.

\begin{figure}[htbp]
    \centering
    \includegraphics[width=1\linewidth]{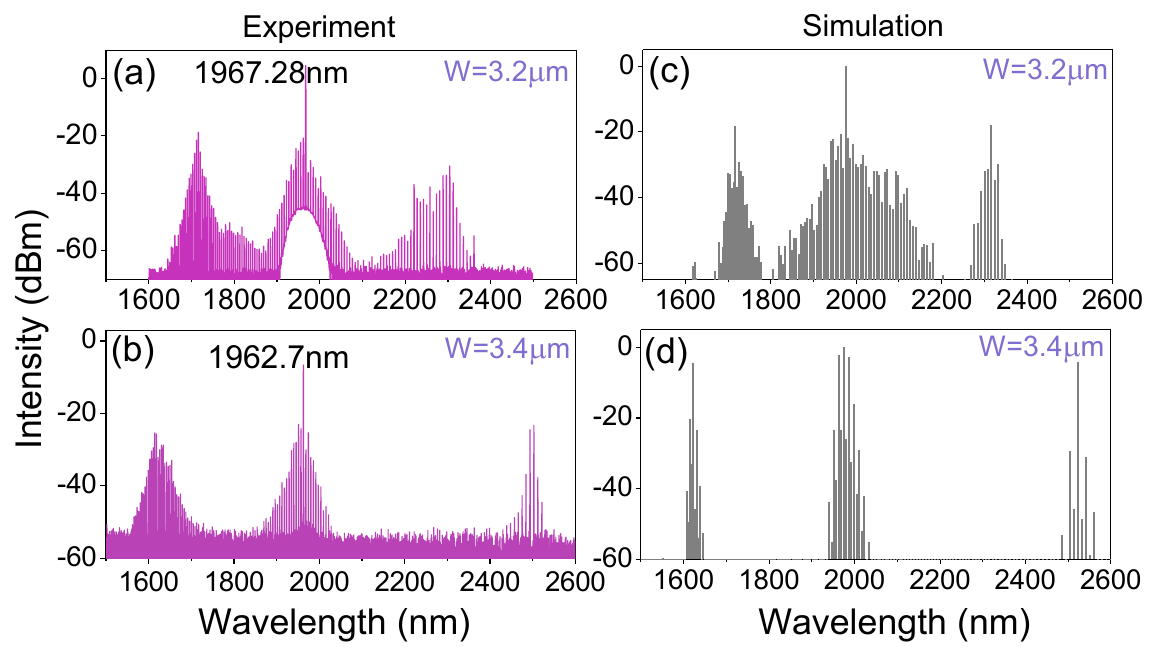}
    \caption{Experimentally measured spectra of comb structures in the microring with R$=$50 $\mu$m and W$=$3.2 $\mu$m (a) and W$=$3.4 $\mu$m (b). (c,d) show the corresponding simulated comb spectra under on-chip pump power of 300 mW.}
    \label{fig.6}
\end{figure}

 In conclusion, we have realized widely-frequency-shifted (from $\sim$45.5 to $\sim$65.5 THz) OPO sources in integrated crystalline AlN microrings pumped at $\sim2 \mu$m, with sidebands extending to the telecom band and the mid-IR region. The experimental results agree well with the numerical simulations. The dispersion engineering via tailoring the microring width modifies the OPO phase-matching condition and gives rise to frequency tuning of the OPO sidebands with offset up to 10-THz. In addition, the OPO sideband frequencies can also be fine-tuned in the 1-THz and 0.1-THz scales by shifting the pump wavelength and applying temperature control, respectively. We believe such wide-spanning on-chip OPO sources with flexible wavelength tunability would enable new capabilities for spectroscopy sensing and nonlinear frequency conversion based on integrated nanophotonic circuits. 

\section*{Acknowledgement}
Yulong Tang acknowledges China Scholarship Council for the visiting research in Yale University. 

\section*{Funding Information}
This work is supported by DARPA SCOUT (W31P4Q-15-1-0006). H.X. Tang acknowledges partial support from DARPA's ACES programs as part of the Draper-NIST collaboration (HR0011-16-C-0118), an AFOSR MURI grant (FA9550-15-1-0029), a LPS/ARO grant (W911NF-14-1-0563), a NSF EFRI grant (EFMA-1640959) and Packard Foundation.   

\section*{Disclosures}
The authors declare no conflicts of interest.

\bibliography{OL_manu}

\bibliographyfullrefs{OL_manu}

\end{document}